\documentclass[prd,superscriptaddress,amsfonts,amssymb,amsmath,showpacs,twocolumn]{revtex4-2}
\usepackage{bm}
\usepackage{amsfonts}
\usepackage{latexsym}
\usepackage[latin1]{inputenc}
\usepackage{graphicx}
\usepackage{amsmath}
\usepackage{palatino}
\usepackage{mathpazo}
\usepackage{textcomp}
\usepackage{float}
\usepackage{booktabs}
\usepackage{dcolumn}
\usepackage{ragged2e}
\usepackage{hyperref}
\hypersetup{colorlinks,citecolor=blue}
\hypersetup{colorlinks=true,linkcolor=red,filecolor=magenta,    urlcolor=blue}
\usepackage{amsmath}
\usepackage{xcolor}
\usepackage{orcidlink}
\usepackage[caption=false]{subfig}
\usepackage{commath}
\def\jnl@style{\it}
\def\aaref@jnl#1{{\jnl@style#1}}

\def\aaref@jnl#1{{\jnl@style#1}}

\def\aj{\aaref@jnl{AJ}}                   
\def\apj{\aaref@jnl{ApJ}}                 
\def\apjl{\aaref@jnl{ApJ}}                
\def\apjs{\aaref@jnl{ApJS}}               
\def\apss{\aaref@jnl{Ap\&SS}}             
\def\aap{\aaref@jnl{A\&A}}                
\def\aapr{\aaref@jnl{A\&A~Rev.}}          
\def\aaps{\aaref@jnl{A\&AS}}              
\def\mnras{\aaref@jnl{Mon.~Not.~Roy.~Astron.~Soc.}}             
\def\prd{\aaref@jnl{Phys.~Rev.~D}}        
\def\plb{\aaref@jnl{Phys.~Lett.~B}}        
\def\prc{\aaref@jnl{Phys.~Rev.~C}}  
\def\prl{\aaref@jnl{Phys.~Rev.~Lett.}}    
\def\qjras{\aaref@jnl{QJRAS}}             
\def\skytel{\aaref@jnl{S\&T}}             
\def\ssr{\aaref@jnl{Space~Sci.~Rev.}}     
\def\zap{\aaref@jnl{ZAp}}                 
\def\nat{\aaref@jnl{Nature}}              
\def\aplett{\aaref@jnl{Astrophys.~Lett.}} 
\def\apspr{\aaref@jnl{Astrophys.~Space~Phys.~Res.}} 
\def\physrep{\aaref@jnl{Phys.~Rep.}}      
\def\physscr{\aaref@jnl{Phys.~Scr}}       
\def\commat{\aaref@jnl{Comm.~Math.~Phys.}}              
\def\science{\aaref@jnl{Science}}               
\def\cqg{\aaref@jnl{Classical Quant.~Grav.}}            
\def\jpcs{\aaref@jnl{JPCS}}                                     
\def\ijmpd{\aaref@jnl{Int.~J.~Mod.~Phys.~D}}                    
\def\grg{\aaref@jnl{Gen.~Relat.~Gravit.}}               
\def\rpp{\aaref@jnl{Rep.~Prog.~Phys.}}          
\def\npa{\aaref@jnl{Nucl.~Phys.~A}}        
\def\lrr{\aaref@jnl{Living Rev.~Rel.}}                   
\def\jcap{\aaref@jnl{J.~Cosmology Astropart.~Phys.}}    
\def\rmp{\aaref@jnl{Rev.~Mod.~Phys.}}   
\def\epjc{\aaref@jnl{Eur.~Phys.~J.~C}}

\allowdisplaybreaks[1]

\begin{document}
\color{black}       
\title{Electrically charged compact stars with an interacting quark equation of state}

\author{Grigoris Panotopoulos \orcidlink{0000-0002-7647-4072}} \email{grigorios.panotopoulos@tecnico.ulisboa.pt}
\affiliation{Centro de Astrof{\'i}sica e Gravita{\c c}{\~a}o-CENTRA, Instituto Superior T{\'e}cnico-IST, Universidade de Lisboa-UL, Av. Rovisco Pais, 1049-001 Lisboa, Portugal}

\author{Takol Tangphati
\orcidlink{0000-0002-6818-8404}} 
\email{takoltang@gmail.com}
\affiliation{Department of Physics, Faculty of Science, Chulalongkorn University, \\Bangkok 10330, Thailand}

\author{Ayan Banerjee \orcidlink{0000-0003-3422-8233}} 
\email{ayanbanerjeemath@gmail.com}
\affiliation{Astrophysics and Cosmology Research Unit, School of Mathematics, Statistics and Computer Science, University of KwaZulu--Natal, Private Bag X54001, Durban 4000, South Africa}


\date{\today}

\begin{abstract}
We investigate the properties of non-rotating, electrically charged strange quark stars in four-dimensional Einstein-Maxwell theory. For quark matter we adopt the well-motivated quantum chromodynamics (QCD) equation-of-state, while for the charge density we assume it is proportional to the mass density. The system of coupled structure equations are integrated numerically, and we compute the mass, the radius as well as the total electric charge of the stars. Moreover, we perform a detailed numerical study of the effect of electric charge using the interacting quark equation-of-state. We find that stars as heavy as two solar masses can be supported, and that the highest radius and highest mass of the stars increase with the charge density.
\end{abstract}

\maketitle

\section{Introduction}

Over the past few decades, compact objects such as white dwarf, neutron star or black hole are a new paradigm of modern astrophysics aimed to address the fundamental principles of physics under
extreme conditions, for example, high temperature and high density circumstance in the core of compact objects. In particular, neutron stars (NSs) born in the aftermath of a core-collapse supernova explosion
\cite{Woosley,Heger:2002by}, are an excellent probe of nuclear matter in extreme environments and the  strong field regime also. Perhaps the composition and the properties of matter in the interior of these objects is the major unsolved problem in stellar astrophysics. Therefore, it is useful to investigate their 
internal structure and the role played by gravity in the formation of compact objects could be an excellent test for the general relativity.

On the other hand, the observations of pulsars heavier than 2 $M_{\odot}$  \cite{Demorest:2010bx,Antoniadis:2013pzd} together
with recent simultaneous determinations of masses and radii
about $R \lesssim (11 \sim 14)$ Km \cite{Steiner:2010fz} have provided a strong constraint on the 
theoretical construction of the NS equation-of-state (EoS). In spite of many efforts to explore the 
EoS in the inner core at ultra-density (above than nuclear density $\rho \gtrsim 3 \rho_{\text{nuc}} $ where
$\rho_{\text{nuc}} = 2.8 \times 10^{14}$ $g/cm^3$ is the normal nuclear density), the problem remains unsolved \cite{Lattimer:2015nhk}. Because,
these densities are unreachable from present laboratory experiments.

This situation motivates us to investigate the properties of strongly correlated quark matter. It means that these stars are not composed of purely neutron matter, but rather that, there is a  phase transition from nuclear matter to quark matter. Such transitions have increased considerably interest in those highly dense cores of NSs and that some stars, called \textit{quark stars}, contain cores of quark matter. In Ref. \cite{Witten:1984rs,Bodmer:1971we}, it was shown that 
the cold strange quark matter could be absolutely stable and the true ground state of hadronic matter. This form of matter consists of equal amounts of up, down and strange quarks, and a 
small number of electrons to attain the charge neutrality. This structure of the quarks leads to a net positive charge inside the star. In this regard, the  MIT-Bag Model 
has been often used to show the stability of the strange matter \cite{Farhi:1984,Madsen:1993iw,Madsen:1998uh},
where each quarks moves freely in a spherically symmetric well of infinity depth (bag).

 If quark matter exists in the cores of NSs, they ought to be made of chemically equilibrated strange matter,
which requires the presence of electrons inside the stars. Thus, the existence of electrons play 
an important role in the formation of an electric dipole layer on the surfaces of quark stars (QSs), 
which may lead to huge electric fields on the order of $10^18$ V/cm \cite{alcock86:a,alcock88:a}. 
In fact, the effect of electric charge and electric field in a self gravitating object has widely been accepted among 
researcher, see the Refs. \cite{Lemos:2014lza,Arbanil:2017huq} and references therein. However, the self gravitating star can originally be traced back to Rosseland in 1924 \cite{Rosseland}, where the star is modeled by a ball of hot  ionized gas, did contain a net charge. In \cite{Ray:2003gt}, the inclusion of electric charge in compact stars have studied by assuming the charge distribution is proportional to the mass density. They found that the charge can be as high as $10^{20}$ Coulomb to bring in any change in the mass-radius relation of the star. 

Based on the simplest realistic linear equation of state, Ivanov \cite{Ivanov:2002jy} found 
a noble scheme for charged static spherically symmetric perfect fluid solutions. This fact brings several consequences. 
For example, charged isotropic/anisotropic fluid solutions have been found for linear or non-linear equations of state
\cite{Varela:2010mf,Kumar:2018rlo,Nasim:2018ghs,Thirukkanesh:2008xc,Panotopoulos:2019wsy}. On the other hand, the structure of stars like QSs with a strong electric field has been analyzed in \cite{Arbanil:2015uoa,Malheiro:2011zz},
considering spheres composed of strange matter that follows the MIT bag model EoS. In \cite{Arbanil:2015uoa}, the hydrostatic equilibrium and the stability against radial perturbation has been deeply analyzed 
for charged strange quark stars. More recently, charged strange QSs with anisotropic matter have been fully investigated in 
Ref. \cite{Panotopoulos:2020hkb,Sunzu:2014yva}. 

Motivated by the above discussion our aim of this paper is to find compact charged spheres made of a charged perfect fluid with an interacting quark EoS. In Section \ref{sec2}, the Einstein-Maxwell system
of equations is expressed for static spherically symmetric
spacetime. In \ref{sec3}, we define the boundary conditions and explain the method used for the numerical integration of the equations. In Section \ref{sec4} we present an overview of a QCD motivated EoS. For simplicity, we assume that charge density is proportional to the
energy density. Section \ref{sec5} is devoted to report the general properties of electrically charged QSs such as their mass-radius and the mass-central mass density relation. Discussion and conclusions are reported in \ref{sec6}. We mostly use geometrized units while deriving various equations, which is, $G = c = 1$. We also use the metric signature $(+, -, -, -)$.

\section{General Relativistic Sellar Structure}\label{sec2}

We restrict our attention to static fluid distribution with
spherical symmetry, whose metric is specified by the line element  ($ds^2 = g_{\nu \mu} dx^\nu dx^\mu$, where $\nu, \mu=0,1,2,3)$,
\begin{eqnarray}
ds^2 = e^{\Phi(r)} dt^{2} -e^{\Lambda(r)}dr^{2}- r^{2} d \Omega^2 \, , \label{metr}
\end{eqnarray}
where $d \Omega^2 = d\theta^2 + \sin^2\theta d\vartheta^2 $ is the line element on the unit 2-sphere with the metric functions $\Phi(r)$ and $\lambda(r)$ depending on $r$ alone. 

The stress-energy tensor $T_{\nu}{}^{\mu}$ of the star consists of two terms, the standard term corresponds to the energy momentum tensor of a perfect  fluid source and the electromagnetic term, 
\begin{eqnarray}\label{em}
  T_{\nu}{}^{\mu} &=& (P +\epsilon)u_{\nu} u^{\mu} + 
  P \, \delta_{\nu}{}^{ \mu}
  \nonumber \\
  &&+\frac{1}{4\pi} \left( F^{\mu l} F_{\nu l} +\frac{1}{4 \pi}
    \delta_{\nu}{} ^{\mu} F_{kl} F^{kl} \right) \, ,
\end{eqnarray}
where $\epsilon(r) $ is the energy density, $P(r)$ is the pressure and $u^{\mu}$ is the $4$-velocity.
The perfect fluid matter distribution implies that the flow of matter is adiabatic, no heat  flow, radiation, or 
viscosity is present. The electromagnetic field tensor $F^{\nu \mu}$ associates with   the second term of Eq. (\ref{em}) satisfy the covariant Maxwell equations 
\begin{equation} [(-g)^{1/2} F^{\nu \mu}]_{, \mu} = 4\pi j^{\nu}
  (-g)^{1/2} \, ,
  \label{ecem}
\end{equation}  
where $j^\mu$ is the four-current density. In this study the static spherically symmetric electric field implies all components of the electromagnetic field tensor vanish, except the radial component  of the electric field $F^{01}$ and the last equation is satisfied if $F^{01}= -F^{10}$. For the radial component one obtains using the Eq.\ (\ref{ecem}), 
\begin{equation}
  E(r) = F^{01}(r)=  \frac{1}{r^2} e^{-(\Phi + \Lambda)/2}  4\pi
    \int_{0}^{r}  r'^2 \rho_{ch} e^{ \Lambda /2} dr' \,
  , \label{comp01}
\end{equation}
where $\rho_{ch} = e^{\Phi/2}j^{0}(r)$ is the electric charge distribution inside the star. From last equation we can define the charge of the system as 
\begin{equation}
  q(r) = 4\pi \int_{0}^{r}  r'^2 \rho_{ch} e^{\Lambda /2}  dr' \, ,  
  \label{Q}
\end{equation}
which does not depend on the timelike coordinate $t$, or equivalently
\begin{equation}
  q'(r) = 4 \pi r^2 \rho_{ch} e^{\Lambda /2} \, .
  \label{Q1}
\end{equation}

Finally, the complete structure of energy-momentum tensor Eq.\ (\ref{em}) reads
\begin{small}
\begin{equation}
T_{\nu}{}^{\mu} =\left( \begin{array}{cccc}
 \epsilon + \frac{q^2}{8\pi r^4} & 0 & 0 & 0 \\
0 & -P + \frac{q^2}{8\pi r^4} & 0 & 0 \\
0 & 0 & -P - \frac{q^2}{8\pi r^4}  & 0 \\
0 & 0 & 0 & -P - \frac{q^2}{8\pi r^4}
\end{array} \right) , \label{TEMch}
\end{equation}
\end{small}
where the electric charge is connected to the electric field through the
relation ${q(r)}/{r^2} = E(r)$. Thus, the nonzero components of the Einstein-Maxwell field equations are
\begin{eqnarray}
  e^{-\Lambda}\left(\frac{1}{r} \frac{d\Lambda}{dr} -\frac{1}{r^{2}}\right)
  +\frac{1}{r^{2}} =  8\pi 
  \left( \epsilon + \frac{q^{2}(r)}{8\pi r^4} \right) \, ,  \label{fe1q} \\
  e^{-\Lambda}\left(\frac{1}{r}\frac{d\Phi}{dr}+\frac{1}{r^{2}}
  \right) -\frac{1}{r^{2}}=  8\pi \left( P -
    \frac{q^{2}(r)}{8\pi r^4} \right) \, . \label{fe2q}
\end{eqnarray}
As usual, we define a new quantity $m(r)$ representing the
mass inside the sphere of radius $r$ is given by
\begin{equation}
  e^{-\Lambda(r)} \equiv 1 - \frac{2  m(r)}{ r} +\frac{ q^2(r)}{ r^2 } \, .
  \label{nord}
\end{equation}
Now, replacing Eq. (\ref{nord}) into Eq. (\ref{fe1q}) it gives \cite{bekenstein71:a,felice95:a}
\begin{equation}
  \frac{dm}{dr} = 4\pi r^2 \epsilon
  +\frac{q}{ r}\frac{dq}{dr} \, . \label{dmel}
\end{equation}
The sum of two terms on the right hand side of Eq.\ (\ref{dmel})  corresponds to the mass-energy of the stellar matter and the mass-energy of the electric field carried the electrically charged star. An additional equation is obtained from the contracted
Bianchi identity $\nabla_{\nu} T_{\nu}{}^\mu =0$, which gives
\begin{equation}
  \frac{d\Lambda}{dr} = -\frac{2}{\left(\epsilon+P\right)}\left(\frac{dP}{dr} -\frac{q}{4 \pi r^4}\frac{dq}{dr} \right) \, . \label{bic}
\end{equation}
Finally, replacing Eq. (\ref{Q}) and the conservation equation
(\ref{bic}) into Eq. (\ref{fe2q}), we get 
\begin{eqnarray}
  \frac{dP}{dr}  & = & - \frac{2 \left( m + 4\pi r^3
      \left( P - \frac{q^{2} }{4\pi r^{4} } \right) \right)}{ r^{2}
    \left( 1 - \frac{2 m}{ r} + \frac{ q^{2}}{r^{2} } \right)}
 \ (P +\epsilon)\nonumber \\ & & +\frac{q}{4 \pi r^4}\frac{dq}{dr} \, .
\label{TOVca}
\end{eqnarray}
which is the modified Tolman-Oppenheimer-Volkoff (TOV) equation to the study of equilibrium  of an electrically charged fluid sphere. For $q \rightarrow 0$, Eq. (\ref{TOVca}) reduces to the known standard TOV equation. 

One now has four equations, (\ref{Q}), (\ref{dmel}) , (\ref{bic}) and (\ref{TOVca}) with six unknown
functions of $r$, i.e., $\Phi$,  $m(r)$, $q(r)$, $\epsilon(r)$, $P(r)$ and $\rho_{ch}(r)$, respectively. Thus, we have two degrees of freedom.  To close the system, one may adopt different strategies with the properties and characteristics of charged fluid sphere (see \cite{Arbanil:2015uoa,Arbanil:2013pua} for more details).
For instance, one may consider an equation of state (EoS) relating the pressure with the energy density of the  fluid. Additionally,  for the electrically charged fluid, it is also needed a relation between the charge distribution and the mass density which we will discuss below.

\section{The exterior vacuum region to the star and the boundary conditions} \label{sec3}

In the final step of this analysis is defining the boundary conditions for the sought solutions.
To guarantee regularity at the origin,
we should establish the following  boundary conditions:
\begin{eqnarray}
&& m(r=0) = 0,~~~q(r=0)=0,~~~~\epsilon(r=0)=\epsilon_c,\\ \nonumber 
&& P(r=0)= P_c,~~~e^{\Lambda(r)}|_{r=0}=1, ~~\text{and}~~P(R)|_{r=R}=0, 
\end{eqnarray}
where $r = R$ is the surface of the star with $P_c$ is the central pressure and $\epsilon_c$ is the central energy density, respectively. Since our aim is to numerically solve the TOV equations, thus the inputs  in the equations is the central pressure $P_c$ where $P_c$ is determined by the chosen EoS once the central density $\epsilon_c$ is fixed. 

The solution is then matched to the exterior Reissner-Nordstr\"{o}m spacetime,
with metric given by
\begin{eqnarray}
ds^2 = F(r) dt^2-\frac{dr^2}{F(r)} - r^{2} d \Omega^2 
\end{eqnarray}
where $F(r)= \left(1-\frac{2 M}{r}+\frac{Q^2}{ r^2}\right)$ with $M$ and $Q$ being respectively the total mass and the total charge of the sphere, respectively. From the continuity of the first fundamental form at the boundary implies that $g_{tt}^{-} = g_{tt}^{+}$ and $g_{rr}^{-} = g_{rr}^{+}$, which yield
\begin{eqnarray}
 \label{eq50}
e^{\Phi(r)}|_{r=R}= F(R) \quad \mbox{and} \quad   e^{-\Lambda(r)}|_{r=R}=F(R),
\end{eqnarray}
with other conditions are $m(R) = M$, $q(R) = Q$, besides $P(R) = 0$.

\section{Equations of State for Quark matter and the charge density
profile} \label{sec4}

\begin{figure}
    \centering
    \includegraphics[width = 7.5cm]{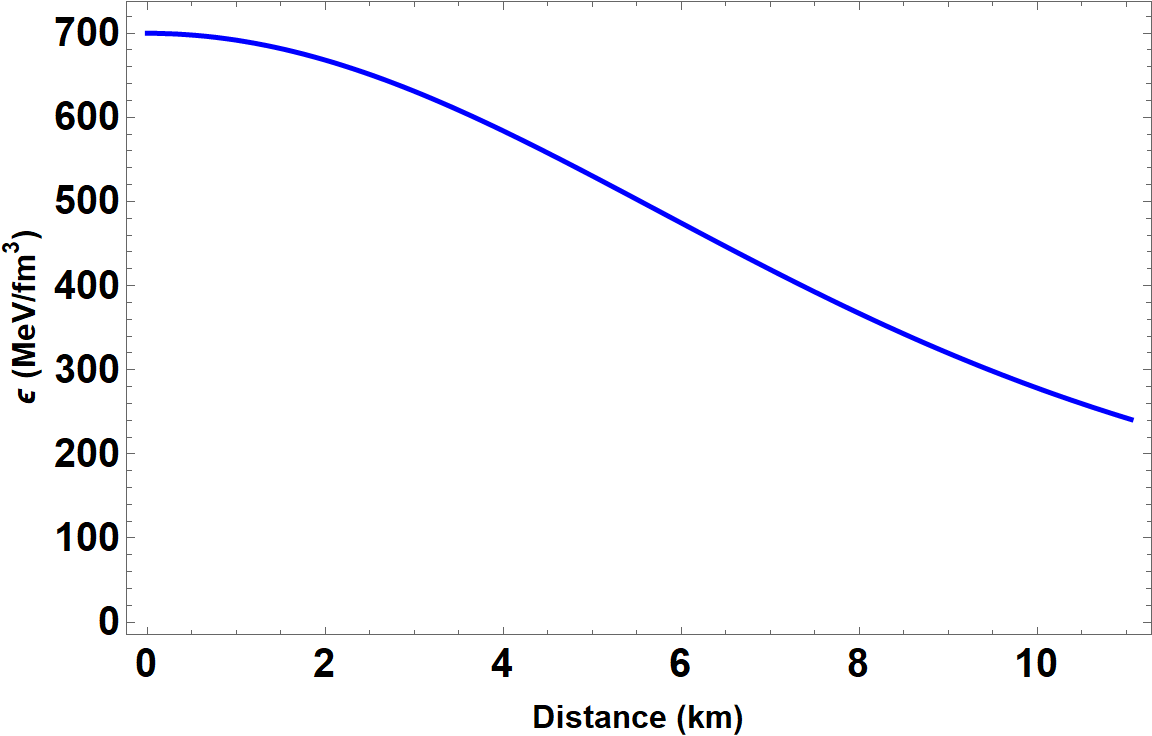}
    \includegraphics[width = 7.5cm]{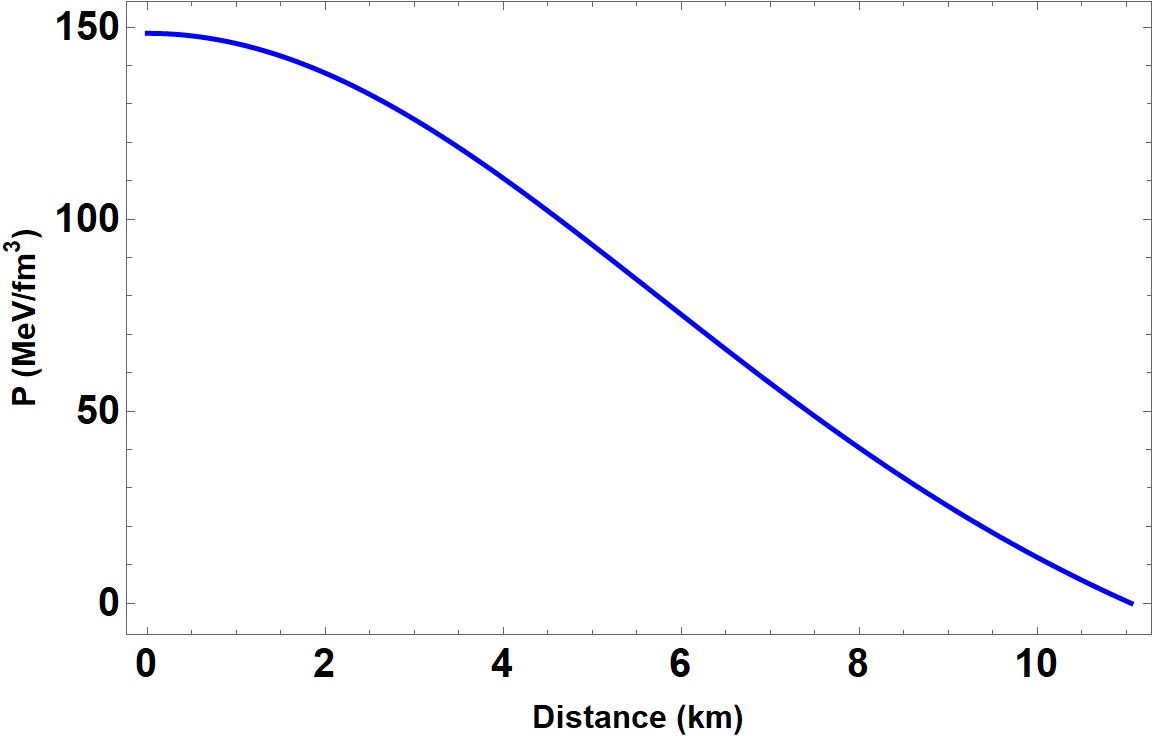}
    \includegraphics[width = 7.5cm]{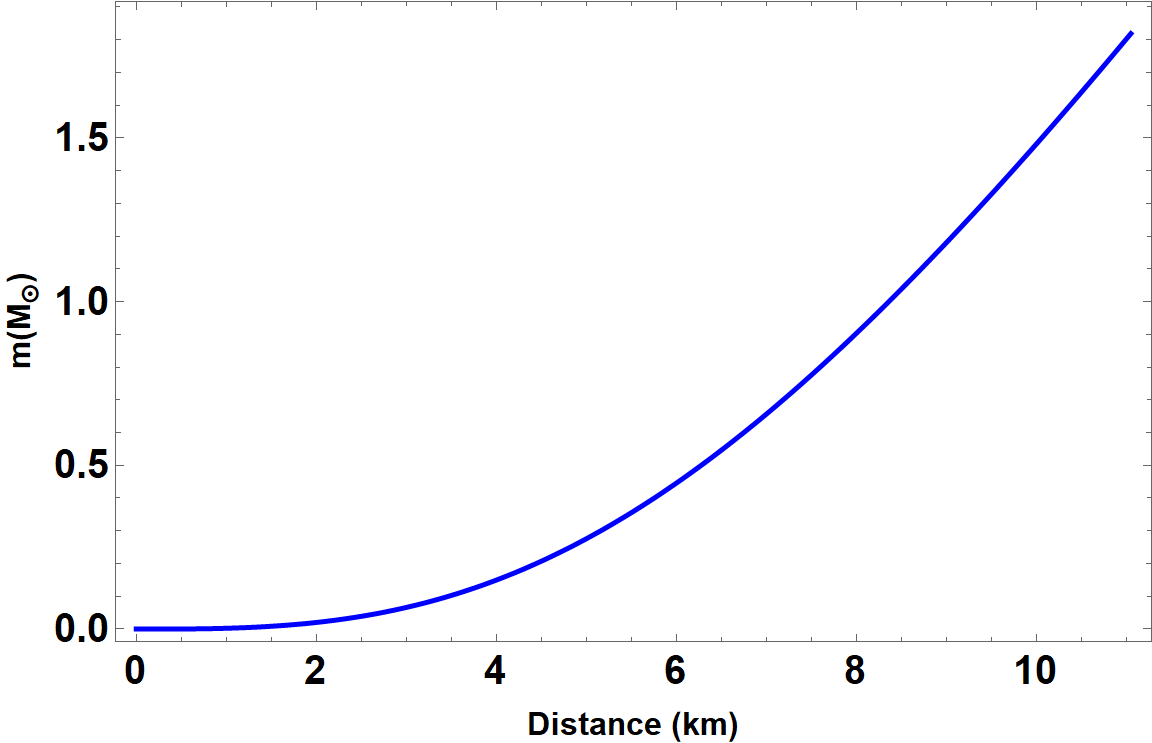}
    \includegraphics[width = 7.5cm]{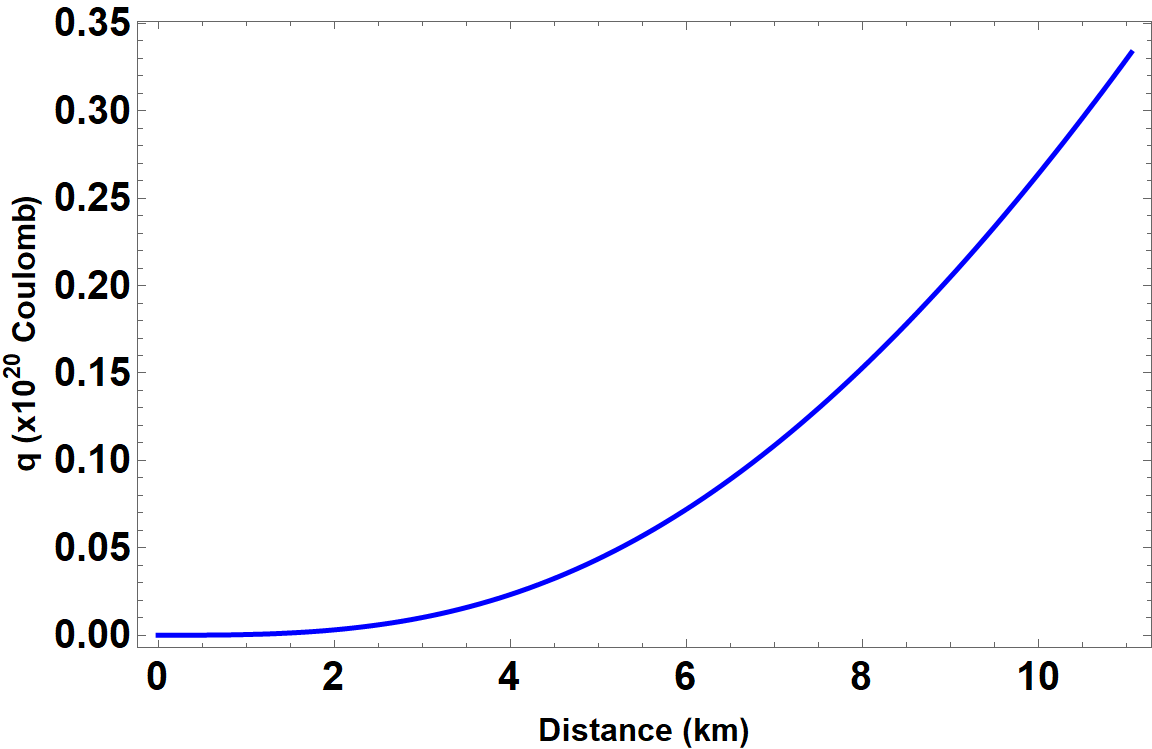}
    \caption{The profiles of mass, energy density, pressure and charge as functions of distance from the center of the star with a set of parameters as $\alpha = 10^{-4}$, $B = 70 \text{ MeV/fm}^3$, $a_4 = 0.9$, $m_s=100 \,{\rm MeV/c^2}$ and $\epsilon_{\text{c}} = 700 \text{MeV/fm}^3$, respectively.}
    \label{example1}
\end{figure}

\subsection{Equation-of-state}

To integrate the structure equations we must specify the sources first.
Since the matter content carries a net electric charge, both an equation-of-state
and an assumption for the charge density is required. As mentioned above, to obtain a solution to the field equations, the matter EoS must be supplied for the charged fluid sphere. The EoS is a thermodynamic equation relating some fundamental quantity, such as the pressure or energy density, and various
(usually intensive) parameters. In general, the structure of NS is composed of two regions, namely the core and the crust. For instance, the core of NSs are composed with supranuclear density, and it could be as high as 10 times the central density
($\epsilon_c = 2.8 \times 10^{14}$ $g/cm^3$) of heavy atomic nuclei.  For this reason, 
 it is believed that high density and relatively low temperature required to produce color superconducting quark matter. 
Thus, it is believed that NSs are the best candidate in the universe where quark matter could be found. 

At present, several models have been proposed in existence of a large variety of color superconducting
states of quark matter at ultra-high densities, see Ref. \cite{Lugones:2002zd,Bogadi:2020sjy,Matsuzaki:2007kg} for details.
To describe matter in the interior of QSs, we assume the EoS that consists of homogeneous and unpaired,  overall electrically neutral, 3-flavor interacting quark matter \cite{Flores:2017kte}. Within this theory, one can describe this phase using the simple thermodynamic Bag model   EoS \cite{Alford:2004pf} with $\mathcal{O}$ $(m_s^4)$ corrections. In this set up several authors have shown that interacting parameter $a_4$ affects significantly the mass-radius relationship of strange stars, allowing for large maximum masses \cite{Becerra-Vergara:2019uzm,Banerjee:2020dad,Panotopoulos:2021sbf}. The QCD motivated EOS can be addressed via explicit expressions for energy density
($\epsilon$) and pressure ($P$) \cite{Becerra-Vergara:2019uzm}
\begin{eqnarray} \label{Prad1}
&& P = \dfrac{1}{3}\left(\epsilon-4B\right)-\dfrac{m_{s}^{2}}{3\pi}\sqrt{\dfrac{\epsilon-B}{a_4}} \nonumber\\
&& +\dfrac{m_{s}^{4}}{12\pi^{2}}\left[1-\dfrac{1}{a_4}+3\ln\left(\dfrac{8\pi}{3m_{s}^{2}}\sqrt{\dfrac{\epsilon-B}{a_4}}\right)\right],
\end{eqnarray}
where $\epsilon$ is the energy density of homogeneously distributed quark matter (also to $\mathcal{O}$ $(m_s^4)$ in the Bag model). For the purpose of the present analysis, following Beringer \textit{et al} \cite{Beringer:2012}, we choose the  strange quark mass is $m_{s}$ to be $100 \,{\rm MeV/c^2}$,  whereas the  accepted values of Bag constant $B$ lies within the range of  $57 \leq B \leq 92$ MeV/fm$^3$, see Ref. \cite{Burgio:2018mcr,Blaschke:2018mqw}. Also, the parameter $a_4$ comes from the QCD corrections on the pressure of the quark-free Fermi sea; is directly related with the  mass-radius relations of QSs. 

\begin{figure}
    \centering
    \includegraphics[width = 7.5cm]{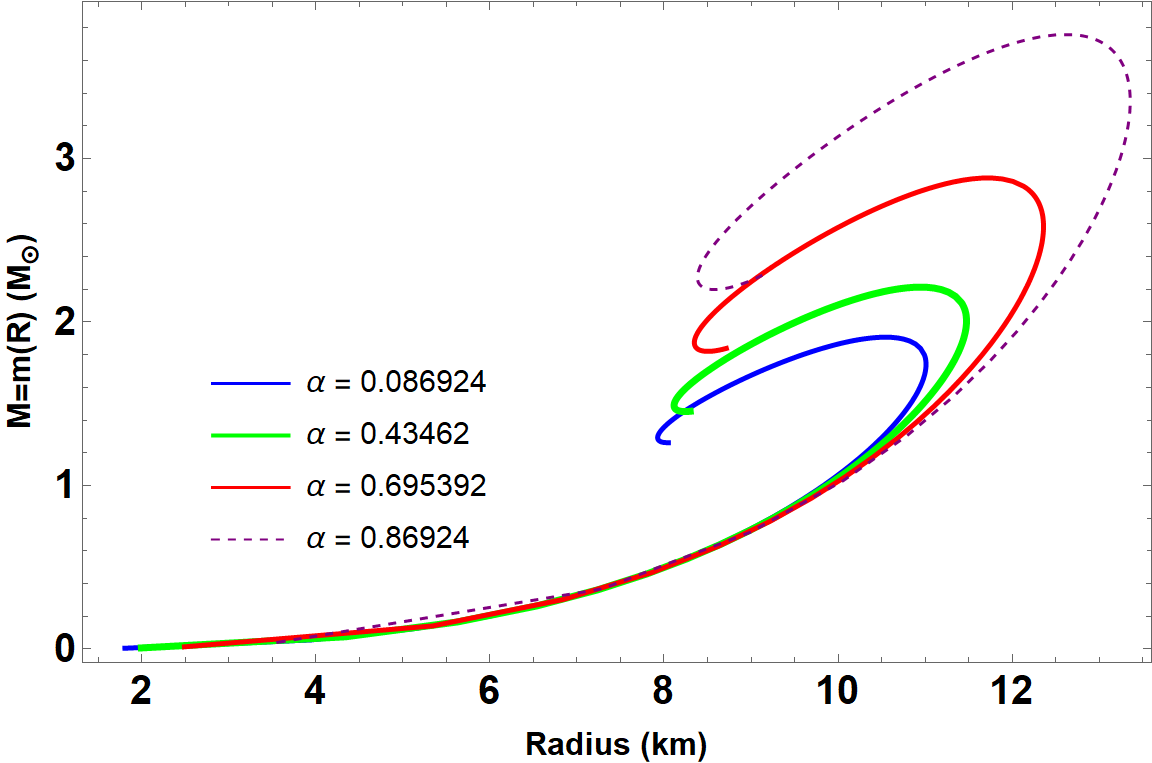}
    \includegraphics[width = 7.5cm]{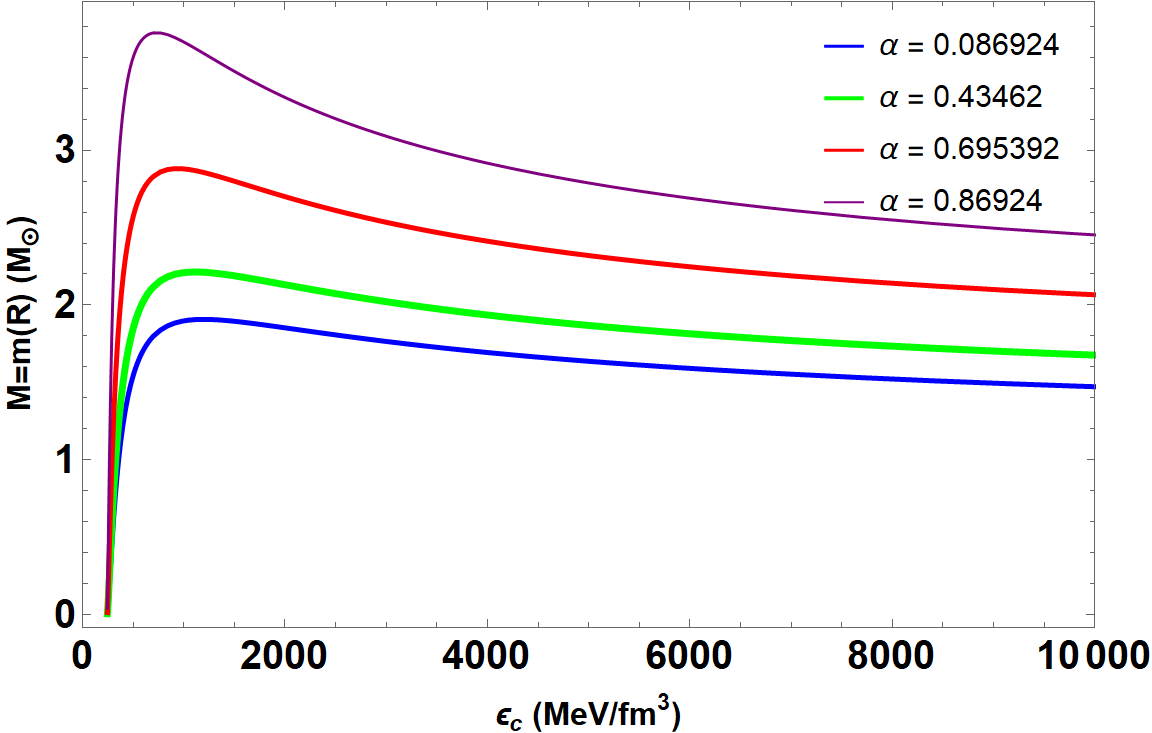}
    \includegraphics[width = 7.5cm]{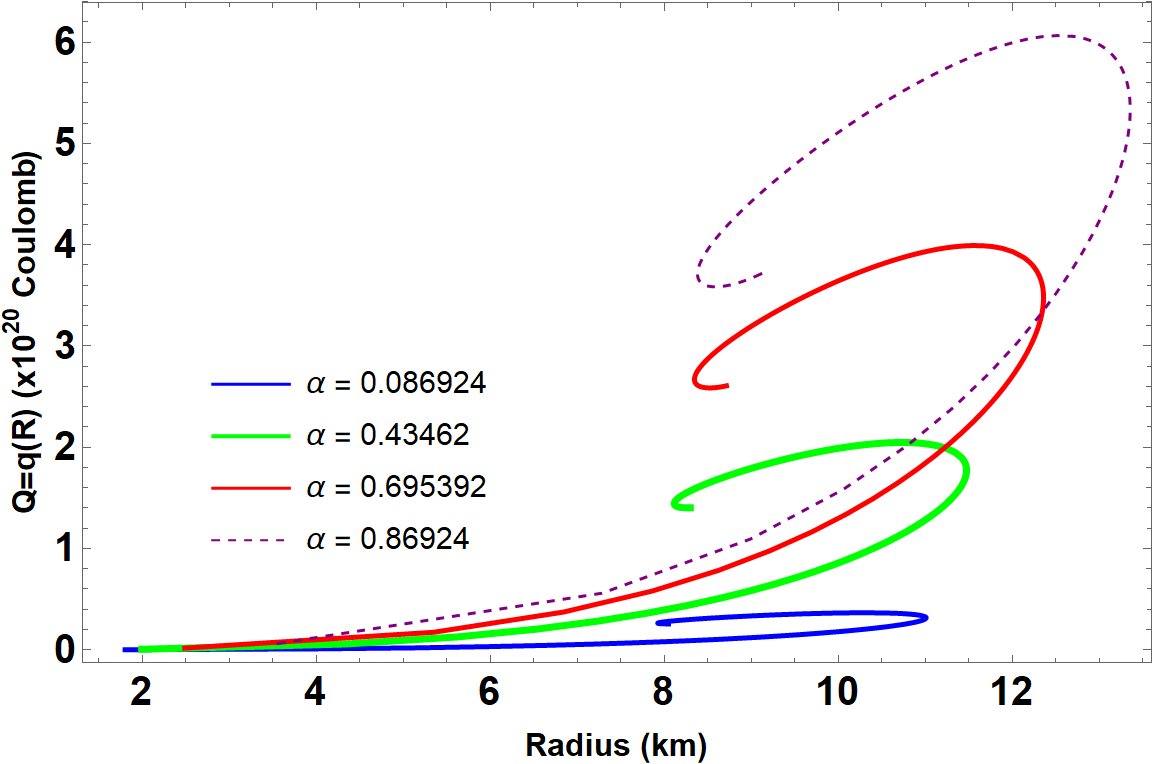}
    \includegraphics[width = 7.5cm]{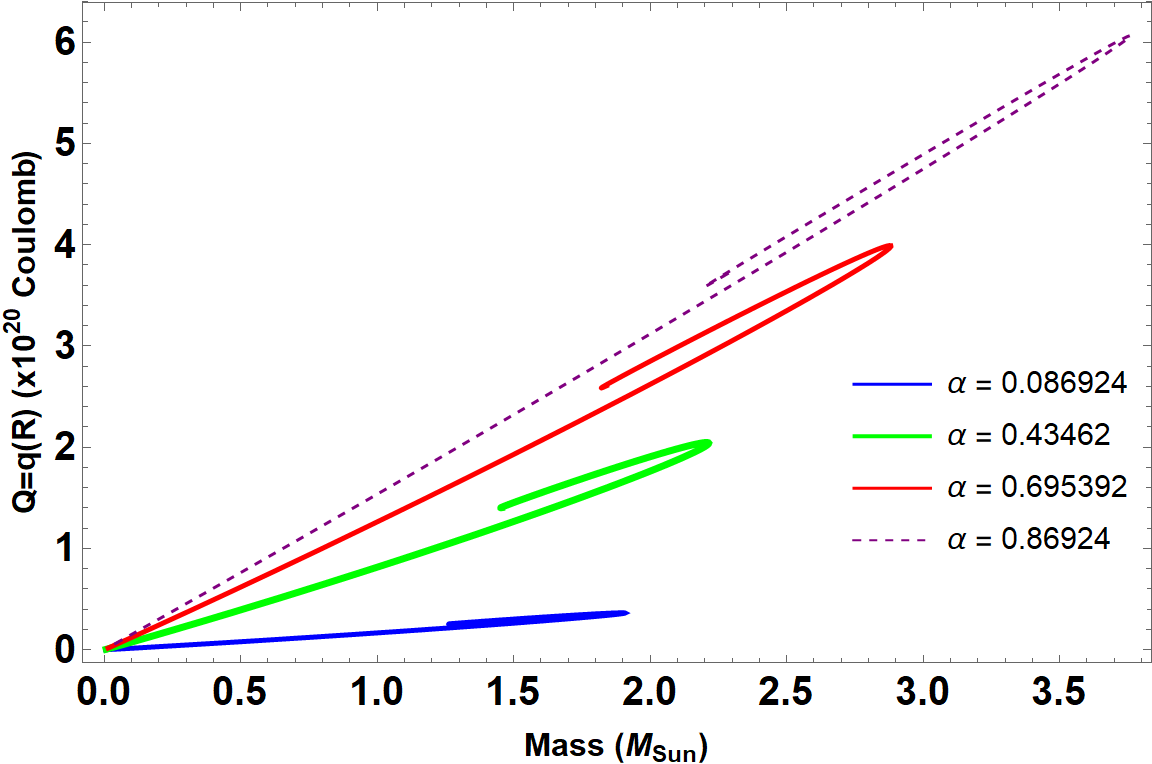}
    \caption{The profiles of mass-radius, mass-energy density, charge-mass, charge-radius with a set of parameters as $B = 70 \text{ MeV/fm}^3$, $a_4 = 0.7$, $m_s=100 \,{~\rm MeV/c^2}$ and $\alpha$ varying from $0.086924$ to $0.86924$.}
    \label{Profiles_alpha}
\end{figure}

\subsection{The charge density relation}

 Since the matter content of the stars carries a net electric charge, we need to specify the charge density as well. Following previous works, for instance \cite{Ray:2003gt,Arbanil:2013pua}, we shall assume in the following that the charge density is proportional to the energy density, i.e.
\begin{equation}
\rho_{ch} = \alpha \: \epsilon \, ,
\end{equation}
with $\alpha$ being the charge fraction. This is a reasonable assumption to make, since more material is expected to bring along a higher amount of electric charge. Clearly
\begin{equation}
\frac{Q}{M} = \frac{\rho_{ch}}{\epsilon} = \alpha,
\end{equation}
and since for the RN solution $Q \leq M$, the charge fraction should take values in the range $0 \leq \alpha \leq 1$.

\begin{figure}
    \centering
    \includegraphics[width = 7.5cm]{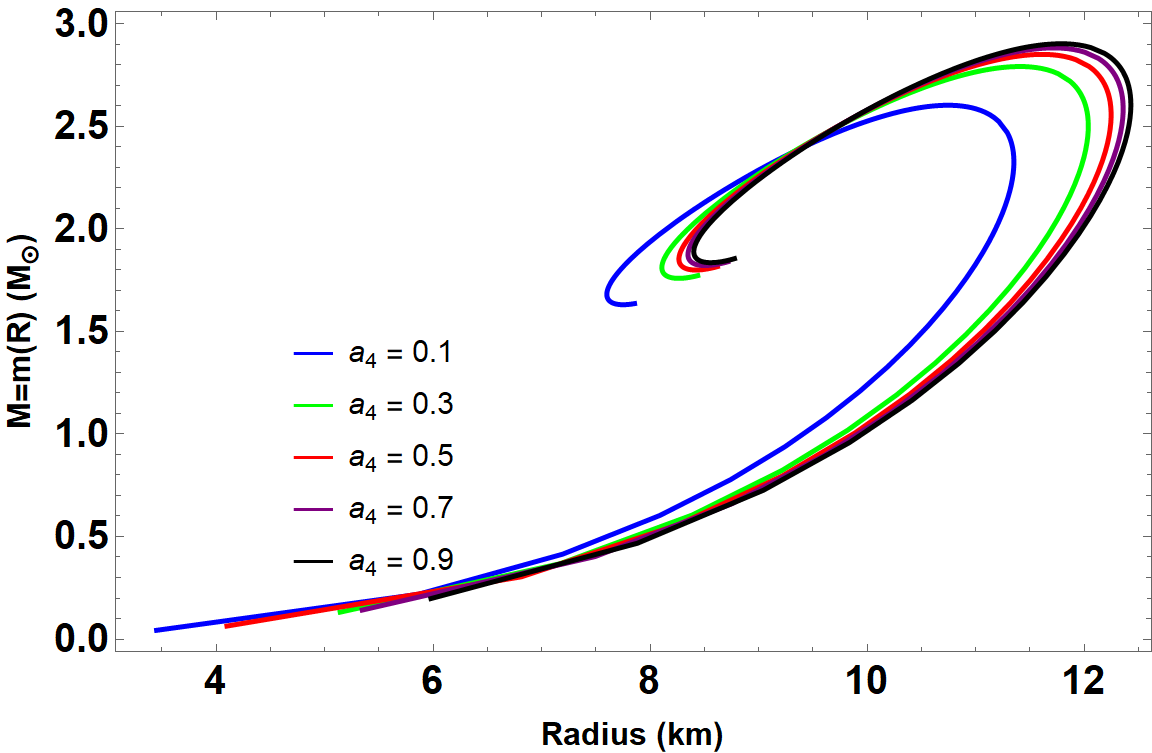}
    \includegraphics[width = 7.7cm]{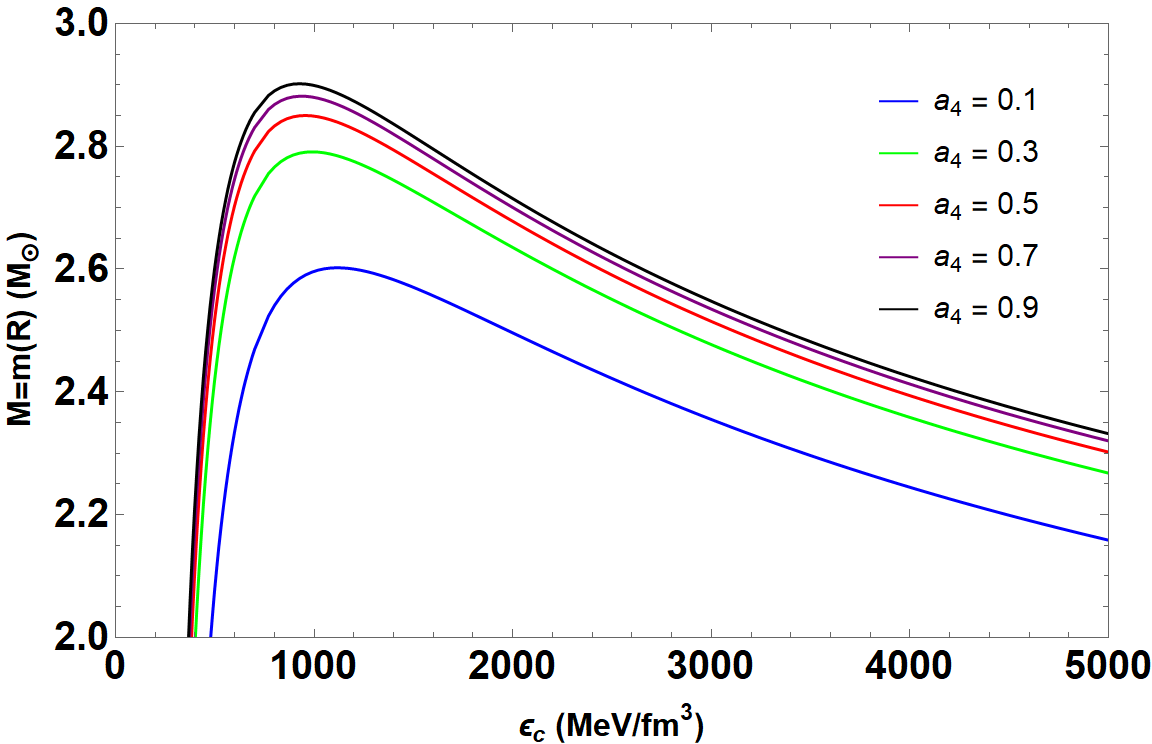}
    \includegraphics[width = 7.5cm]{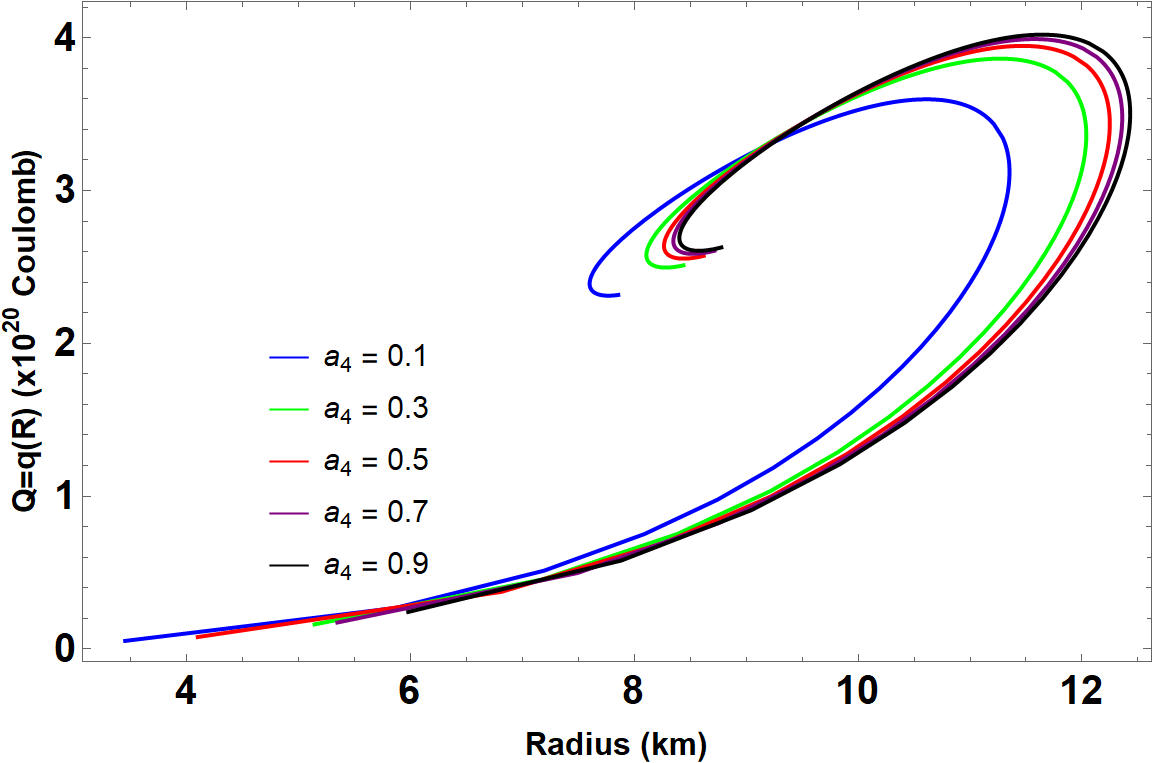}
    \includegraphics[width = 7.5cm]{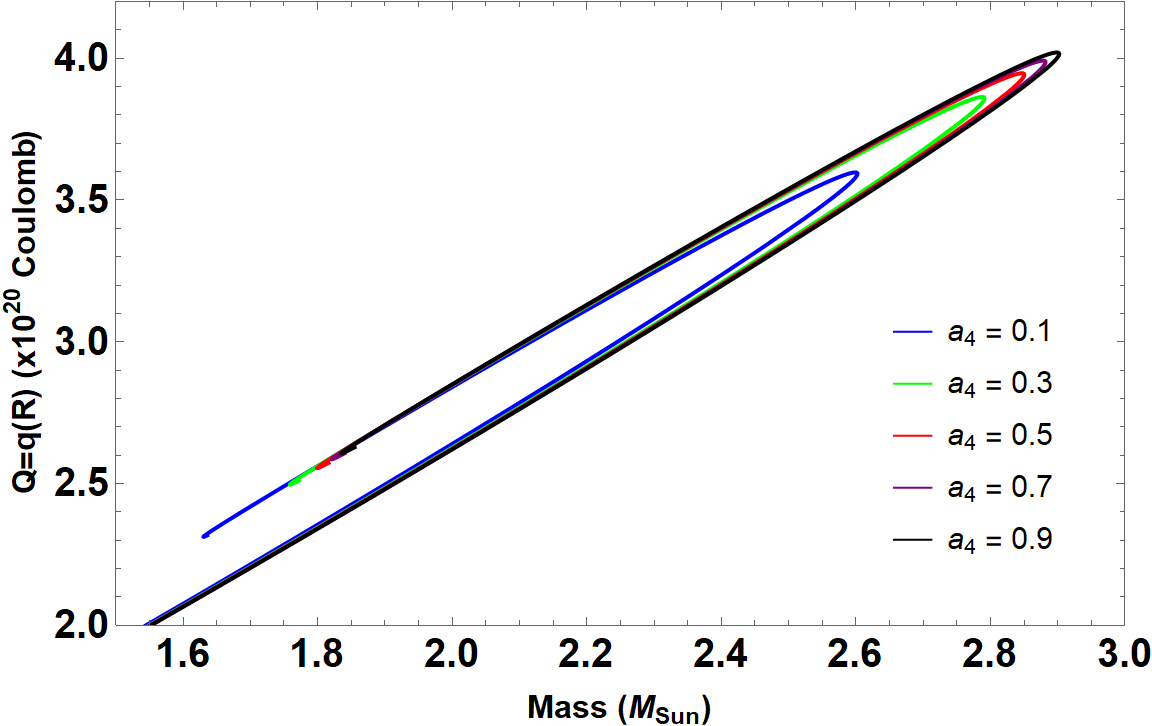}
    \caption{The profiles of mass-radius, mass-energy density, charge-mass, charge-radius with a set of parameters as $B = 70 \text{ MeV/fm}^3$, $\alpha = 0.695392$, $m_s=100 \,{\rm MeV/c^2}$ and $a_4$ varying from $0.1$ to $0.9$.}
    \label{Profiles_a4}
\end{figure}

\section{Numerical results}\label{sec5}

\subsection{Sequences of Electrically Charged Strange Stars}

In our work there are four parameters in total, namely the charge fraction, $\alpha$, the bag constant, $B$, the mass of the s quark, $m_s$ and the interacting parameter, $a_4$. The latter is dimensionless, whereas all the other parameters are dimensionful. The mass of the stars is measured in solar masses, $M_{\odot}$, the radius of the stars in $km$, the total electric charge in $C$, the pressure, the energy density and the bag constant in $\text{MeV/fm}^3$, and finally 
the mass of the s quark in ${\rm MeV/c^2}$. In the following we set $m_s = 100 \, {\rm MeV/c^2}$, $B = 70$ $\text{MeV/fm}^3$, and we vary $a_4,\alpha$ in the range $a_4 = (0.1-0.9)$ and $\alpha = (10^{-4}-10^{-3})$. 

In the four panels of Fig.~\ref{example1}, we show a particular interior solution for a given central 
energy density, $\epsilon_c=700$ $\text{MeV/fm}^3$, and for $a_4=0.9$ and $\alpha=10^{-4}$, respectively. The central value of the pressure is determined by using the EoS. From top to bottom we show the energy density, $\epsilon(r)$, the pressure, $P(r)$, the mass function, $m(r)$, and the electric charge function, $q(r)$, as a function of the radial coordinate $r$. The mass function and the charge function increase monotonically with $r$ starting from zero at the center of the star, while the pressure 
and the energy density decreases monotonically with $r$, starting from their central values. The pressure vanishes at the 
surface of the star, whereas the energy density acquires a non-vanishing surface value $\epsilon_s$.

Next, in Fig.~\ref{Profiles_alpha} we show the impact of $\alpha$ on the profiles $M-R$ and $Q-R$ setting $a_4=0.7$ and varying $\alpha$
from 0.086924 to 0.86924. In each panel we have plotted four curves as follows: Blue color corresponds to $\alpha=0.086924$, 
green color to $\alpha=0.43462$, red color to $\alpha=0.695392$ and purple color corresponds to $\alpha=0.86924$. Panels
1 and 3 show the mass of the star, $M$, and the electric charge of the star, $Q$, versus the radius of the star, $R$,
respectively. In panel 2 we show the mass of the star as a function of the central energy density, and finally in panel
4 we show the total electric charge of the star versus its mass. 

In panel 2 the mass of the star reaches a maximum value at $\epsilon_c^*$, which grows with the charge fraction, in agreement with what is shown in the $M-R$ profile. Moreover, according to the  Harrison-Zeldovich-Novikov criterion \cite{harrison,ZN}
\begin{equation}
\frac{dM}{d \epsilon_c} > 0  \; \; \; \rightarrow \textrm{stable configuration}
\end{equation}
\begin{equation}
\frac{dM}{d \epsilon_c} < 0  \; \; \; \rightarrow \textrm{unstable configuration}
\end{equation}
only the first part of the curve, before the maximum value, corresponds to a stable configuration. Therefore, the point at the extremum of the curve separates the stable from the unstable configuration.

The $Q-R$ profile is similar qualitatively to the $M-R$
profile, which exhibits the typical behaviour observed in strange quark stars with a highest radius and a highest mass.
For a given radius, higher charge fraction, $\alpha$, implies both a higher mass and a higher electric charge. Both the 
highest radius and the highest mass increase with $\alpha$. The same holds for the highest electric charge. The highest values as well as the corresponding central energy density are reported in Table \ref{tab:table1} for $\alpha$ variations.

Finally, Fig.~\ref{Profiles_a4} is similar to the  Fig.~\ref{Profiles_alpha}, but this time we show the impact of $a_4$ on the properties of the stars
setting $\alpha=0.695392$ and varying $a_4$ from 0.1 to 0.9. In each panel we have plotted 5 curves as follows:
Blue color corresponds to $a_4=0.1$, green color to $a_4=0.3$, red color to $a_4=0.5$, purple to $a_4=0.7$ and
black color corresponds to $a_4=0.9$. The shape of the profiles $M-R$ and $Q-R$ is qualitatively similar to
those of Fig.~\ref{Profiles_alpha}. The highest mass, highest radius and highest electric charge increase with $a_4$, while at the
same time we observe that for a given mass or a given electric charge, a larger value of $a_4$ implies a larger
value for the radius of the star. The highest values as well as the corresponding central energy density are reported in Table \ref{tab:table2} for $a_4$ variations.
Once again, in panel 2 i) the mass of the star reaches a maximum value at some $\epsilon_c^*$, which grows with $a_4$, in agreement with what is shown in the $M-R$ profile, and ii) the maximum
of the curve separates the stable from the unstable configuration according to the 
Harrison-Zeldovich-Novikov criterion mentioned before.


\begin{table}[h]
  \caption{\label{tab:table1} Properties of electrically charged quark stars: Shown are the highest values $M,R,Q,E$ for stars assuming $a_4 = 0.7$ and varying $\alpha$. The quantities $M$, $R$ and $\epsilon_c$ denote maximum gravitational masses, radii, and the corresponding central energy densities in MeV/fm$^3$, respectively. The stars carry given electric charges, $Q$ ~ ($\times 10^{20}C$), which gives rise to stellar surface electric fields $E$ ~ ($\times 10^{22}$ V/m).}
\begin{ruledtabular}
\begin{tabular}{cccccc}
$\alpha$ & $M$~ ($M_{\odot}$) & $R$~ (km) & $Q$ & $E$ & $\epsilon_c$\\
\hline
    0.0869 &  1.91  & 10.55 & 0.363 & 0.294 & 1190 \\
    0.4350 &  2.21  & 10.94 & 2.040 & 1.530 & 1120 \\
    0.6950 &  2.88  & 11.76 & 3.980 & 2.590 & 910 \\
    0.8690 &  3.76  & 12.70 & 6.060 & 3.380 & 700 \\
\end{tabular}
\end{ruledtabular}
\end{table}

\begin{table}[h]
  \caption{\label{tab:table2} The obtained values of the maximum-mass, radius and their corresponding central energy density for several different values of $a_4$ with a fixed value of charge fraction $\alpha = 0.695$. The other notations are the same as in Table \ref{tab:table1}.}
\begin{ruledtabular}
\begin{tabular}{cccccc}
$a_4$ & $M$~ ($M_{\odot}$) & $R$~ (km) &$Q$ & E & $\epsilon_c$\\
\hline
    0.1 &  2.60  & 10.71 & 3.59 & 2.82 & 1112 \\
    0.3 &  2.79  & 11.36 & 3.86 & 2.69 & 988 \\
    0.5 &  2.85  & 11.51 & 3.95 & 2.68 & 956 \\
    0.7 &  2.88  & 11.79 & 3.98 & 2.58 & 925 \\
    0.9 &  2.90  & 11.84 & 4.01 & 2.57 & 925 \\
\end{tabular}
\end{ruledtabular}
\end{table}

\section{Conclusions}\label{sec6}

In summary, in the present work we have studied relativistic stars with a net electric charge in the Einstein-Maxwell theory. In particular, we have investigated the properties of non-rotating strange quark stars adopting the interacting equation-of-state, and assuming that the charge density is proportional to the mass density. First we presented the structure equations supplemented by the appropriate boundary conditions, both at the center and at the surface of the stars, demanding that at the surface the interior solution matches the exterior Reissner-Nordstr{\"o}m geometry. Next, we integrated the system of coupled equations (TOV plus Maxwell) to obtain i) the interior solution describing hydrostatic equilibrium of the stars, and ii) the mass-to-radius and charge-to-radius profiles. In the model analyzed here there are in total four parameters. In the numerical analysis we fixed the numerical values of the bag constant as well as the mass of the s quark, and we varied i) the interacting parameter, $a_4$, from 0.1 up to 0.9, and ii) the charge fraction, $\alpha$, from 0.0869 up to 0.869. Our findings show that the mass-to radius profile exhibits the typical shape characterizing quark matter with a maximum radius and a maximum mass, and that the EoS assumed here is capable of supporting stars with a mass at two solar masses. What is more, we find that the highest radii, masses and total electric charges (reported in Tables I and II) increase both with the interacting parameter and the charge fraction. In the case of $\alpha$ variations, the $M-R$ profiles are indistinguishable below $R \approx 9.5~km$, while in the case of 
$a_4$ variations, the profiles are indistinguishable below $R \approx 6~km$.

\section*{Acknowledgments}

The author G.~P. thanks the Fun\-da\c c\~ao para a Ci\^encia e Tecnologia (FCT), Portugal, for the financial support to the Center for Astrophysics and Gravitation-CENTRA, Instituto Superior T\'ecnico, Universidade de Lisboa, through the Project No.~UIDB/00099/2020 and No.~PTDC/FIS-AST/28920/2017. The author T.~T. thanks the Science Achievement Scholarship of Thailand (SAST) for the financial support.





\end{document}